\documentclass{aa}

\usepackage{graphicx}
\usepackage{txfonts}
\usepackage{natbib}
\usepackage{bm}
\usepackage{wasysym}
          
\usepackage{epsfig}          
\usepackage{color}           
\usepackage{hyperref}
\usepackage{breakurl}        
\usepackage[utf8]{inputenc}
\usepackage{float}
\usepackage[english]{babel}

\newcommand{\dg}{$^{\circ}$}

\newcommand{\rsun}{$R_{\odot}$}
\newcommand{\fe}{Fe~{\sc{xiv}}}
\newcommand{\ovi}{O~{\sc{vi}}}
\newcommand{\hi}{H~{\sc{i}} Ly$\alpha$}
\newcommand{\sixii}{Si~{\sc{xii}}}
\usepackage[normalem]{ulem}

\begin{document}

\title{Differential rotation of the solar corona: \\ A new data-adaptive multiwavelength approach}

\author{S. Mancuso\inst{1} \and S. Giordano\inst{1} \and D. Barghini \inst{1,2} \and D. Telloni\inst{1}}
\institute{
INAF -- Osservatorio Astrofisico di Torino, via Osservatorio 20, 10025 Pino Torinese, Italy \and
Dipartimento di Fisica, Universit\`a degli Studi di Torino, via Pietro Giuria 1, 10125 Torino, Italy\\ \email{salvatore.mancuso@inaf.it}}
\date{Received / Accepted}

\abstract
{
The characterization of the differential rotation of the extended corona is still lacking conclusive results about the actual rotation rate profiles, and it is also expected to vary along the solar cycle. 
While some studies supported the quasi-rigidity of coronal rotation, others have found evidence of differential rotation to occur.
}
{
For the purpose of investigating the differential rotation of the solar corona, we analyzed ultraviolet (UV) spectral line observations acquired on both the east and west limbs at 1.7 \rsun\ by the Ultraviolet Coronagraph Spectrometer (UVCS) on-board the {\it Solar and Heliospheric Observatory} ({\it SOHO}) during the solar minimum preceding solar cycle 23.
To obtain a reliable and statistically robust picture of the rotational profile, we used a set of simultaneous 400-day long spectral line intensities of five different spectral lines: \ovi\ 1032 \AA, \ovi\ 1037 \AA, \sixii\ 499 \AA, \sixii\ 521 \AA, and \hi\ 1216 \AA, which are routinely observed by UVCS.
}
{
The data were analyzed by means of two different techniques: the generalized Lomb-Scargle periodogram (GLS) and a multivariate data-adaptive technique called multichannel singular spectrum analysis (MSSA). 
Among many other positive outcomes, this latter method is unique in its ability to recognize common oscillatory modes between the five time series observed at both limbs.
}
{
The latitudinal rotation profile obtained in this work emphasizes that the low-latitude region of the UV corona (about $\pm 20$\dg\ from the solar equator) exhibits differential rotation, while the higher-latitude structures do rotate quasi-rigidly. 
Moreover, in contrast to previous results obtained using only \ovi\ 1032 \AA\ data over a 365-day time interval during solar minimum activity, the alleged north-south rotational asymmetry of the UV corona, if existent, is much less pronounced.
}
{
The differential rotation rate of the solar corona as evinced at low-latitudes is consistent with the rotational profile of the near-surface convective zone of the Sun, suggesting that the rotation of the corona at 1.7 \rsun\ is linked to intermediate-scale magnetic bipole structures anchored near 0.99 \rsun. 
The quasi-rigid rotation rate found at mid and high latitudes is instead attributed to the influence of large-scale coronal structures linked to the rigidly rotating coronal holes.
We further suggest that the methodology presented in this paper could represent a milestone for future investigations on differential rotation rates when dealing with simultaneous multiwavelength data.
}

\titlerunning{Differential rotation of the solar corona: a new data-adaptive multiwavelength approach}
\authorrunning{Mancuso et al.}

\maketitle

\section{Introduction}

A general understanding and agreement have now been established concerning the rotation rate as a function of latitude and depth in the solar interior, as well as in the photospheric and chromospheric layers of the Sun. 
However, such a consensus is still lacking regarding the actual profile of the differential rotation of the global corona along the solar cycle.
Previous research has shown that measuring coronal rotation rates with different tracers and techniques often leads to incompatible results.
On one hand, some studies have suggested that large-scale coronal structures rotate more rigidly than the underlying photosphere (e.g., \citealt{Timothy1975,Bohlin1977,Hansen1969,Parker1982,Fisher1984,Weber1999,Lewis1999,Giordano2008}). 
On the other hand, other studies have proposed that the corona exhibits a significant degree of differential rotation (e.g., \citealt{Shelke1985,Obridko1989,Navarro1994,Insley1995,Brajsa2004,Karachik2006,Chandra2010,Wohl2010}).
This discrepancy can be at least partially explained if we consider that the extended corona is optically thin. 
As a consequence, the presence of prominent rotating tracer features in the coronal environment is not clearly discernible, both in terms of the spatial and temporal extent.
Moreover, there are major problems concerning projection effects at high heliocentric distances and high heliolatitudes.

The actual rotation rate of the solar corona as a function of heliolatitude thus remains largely undetermined, even now, after decades of effort, as it depends both on the types of observational data and the specific techniques used for their analysis.
It is particularly difficult to explain how the corona can manifest rigid rotation when the photospheric magnetic fields with which the coronal structures are associated actually manifest a monotonously decreasing rotation rate all the way from the equator to the solar poles.
In this respect, it is clear that for any given tracer that may be used to probe the coronal rotation, a distinction must be made among short-lived features (such as coronal bright points) and longer-lived structures (such as coronal streamers or holes).
Moreover, the rotation rate of the corona is certainly inherently coupled with the rotation rate of both the interior and the surface of the Sun.
As a consequence, small-scale coronal structures are presumably doomed to be linked to correspondingly small-scale photospheric magnetic fields and are thus expected to rotate differentially as well.
On the other hand, large-scale coronal structures, probably rooted much deeper (but how much remains to be determined) in the interior of the Sun, could manifest a different rotational profile (but we do not know why it is quasi-rigid) with respect to heliolatitude according to the depth at which the roots are attached in the interior of the Sun (see, e.g., \citealt{Hiremath2013} for a discussion).
Another plausible explanation is that the structures in the corona are only sensitive to large-scale magnetic-field variations and low multipole moments in the underlying photosphere (\citealt{Wang1988}), so a gradual transition from rigid to differential rotation should be expected with decreasing distance from the photosphere and with increasing spatial resolution of the data.
Mechanisms such as magnetic interchange reconnection (\citealt{Crooker2002}) at coronal hole boundaries between closed and open fluxes could also contribute to the observed rigid rotation (\citealt{Wang1993}).
However, no convincing interpretation of the quasi-rigid rotation of the corona has been given as yet.

The rotation period of the extended corona can be conveniently measured by using observations of ultraviolet (UV) spectral lines that are sensitive to coronal activity.
This method relies on the presence of extended, stable coronal structures (streamers) often associated with solar active regions or sunspots. 
While rotating with the Sun, these tracers are able to imprint a temporal $\sim 27$-day modulation on the flux of specific UV emission spectral lines, as observed by spectrographs above the limbs of the Sun, thus providing a means to probe the angular velocity within the coronal layers where those spectral lines are formed.
In previous investigations of the UV coronal rotation rate at various heliocentric distances throughout solar cycle 23 (\citealt{Giordano2008}; \citealt{Mancuso2011,Mancuso2012,Mancuso2013}), only a specific spectral line, \ovi\ 1032~\AA, was used because of its brightness and the fact that it represents one of the lines most sensitive to coronal activity within the range of UV spectral lines probed by the Ultraviolet Coronagraph Spectrometer (UVCS; \citealt{Kohl1995}) on-board the {\it Solar and Heliospheric
Observatory} ({\it SOHO}; \citealt{Domingo1995}).
However, especially during the first year of functioning of the UVCS instrument (corresponding to a phase of solar minimum activity), several other prominent coronal lines, such as \ovi\ 1037 \AA, \sixii\ 499 \AA, \sixii\ 521 \AA, and \hi\ 1216 \AA\ were also regularly observed with almost daily cadence at lower and mid heliolatitudes.
In general, UVCS observations of the streamer belt during solar minimum activity have shown that the \hi\ line intensity presents a simple structure, as is also observed in white-light coronagraph images, with maximum brightness centered on the streamer belt plane and more diffused brightness toward its edges.
On the other hand, the intensity of the \ovi\ lines for the same coronal features is generally more structured, often showing a clear bifurcated structure, suggesting the presence of two sub-streamers located outside the streamer-belt plane (e.g., \citealt{Abbo2019}).
\sixii\ lines are instead more sensitive to the high-density and high-temperature plasma associated with localized active regions. 
Therefore, using the extended dataset of multiple spectral lines  acquired in the first year of functioning of UVCS, it is possible, other than to improve the statistics, in principle, to provide a more reliable picture of the rotation rate of the global corona. 
In fact, results obtained through the analysis of a single spectral line might only reflect specific structures of the corona, as discussed above.
The aim of this work is thus to complement and extend the previous analysis of \cite{Giordano2008} on the differential rotation of the UV corona by using a set of intensity time series from five different emission spectral lines, four of which have not been fully analyzed before. 
As a plus, the periodicity analysis of the newly calibrated data was implemented in an original multivariate fashion by adopting an advanced data-adaptive technique that is able to take into proper account the cross-correlation between the signals obtained at the two (east and west) limbs.
We show that the multivariate analysis of the full set of UV data improves both the accuracy and the reliability of the estimate of the coronal rotation period. 
As a matter of fact, the methodology presented in this paper could actually represent a milestone for future investigations on the differential rotation of the Sun when dealing with simultaneous observations of coronal spectral lines.
The plan of this paper is as follows.
In Sect. 2, we describe the observations and the data reduction. 
In Sect. 3, we give a description of the two methods that have been implemented for extracting the information on the periodicity from the data. 
In Sect. 4, we present our analysis and results. 
Finally, after a discussion in Sect. 5, we summarize our conclusions in Sect. 6.

\begin{figure*}
\centering
\includegraphics[width=12cm]{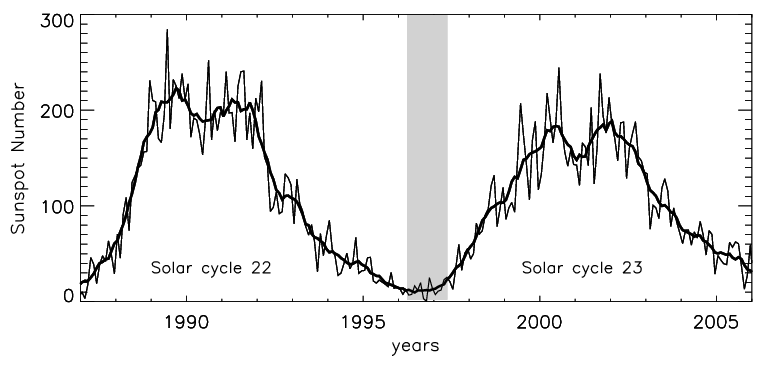}
\caption{
Monthly and smoothed sunspot numbers from 1987 to 2006. 
The shaded region shows the time interval that has been used for the study of the coronal rotation at solar minimum. 
Courtesy of Solar Influences Data Analysis Center (SIDC), Belgium.
}
\label{Fig1}
\end{figure*}

\section{Observations and data analysis}

We focused our investigation on a period of minimum solar activity in which the solar corona was relatively stable due to the reduced frequency of coronal mass ejections (CMEs) and of the low flaring activity.
In particular, we selected a 400-day observation period, from 1996 April 4 to 1997 May 15 (see Fig. 1), that nicely corresponds to the minimum of the sunspot number between solar cycles 22 and 23.
In that time interval, the corona displayed a quasi-dipole axisymmetric global magnetic field configuration and was characterized, as seen in  white-light coronagraphic images, by a stable and wide equatorial streamer belt together with higher-latitude long-lived coronal structures linked to sunspots and solar active regions. 
The data were acquired by {\it SOHO}/UVCS, which is an internally and externally occulted coronagraph consisting of two toric grating spectrometers (channels) for the observation of spectral lines in the UV range. 
During its first years of functioning, UVCS routinely ran daily synoptic observations in which the corona was observed between about 1.4 \rsun\ and 3.5 \rsun\ by moving the 42\arcmin\ entrance slit in a number of different radial locations at eight different position angles (PAs; measured counterclockwise from the north heliographic pole) separated by an angular step of 45\dg.
For this work, we considered data taken from both UVCS channels, that is, the LYA channel (optimized for line measurements of \hi\ 1216~\AA), and the OVI channel (optimized for measurements of \ovi\ 1032~\AA\ and \ovi\ 1037~\AA).
This latter channel was also used for the detection of spectral lines in the 492 \AA\ to 540 \AA\ (second-order) spectral range, thus allowing the simultaneous and cospatial measurements of the line profiles of \sixii\ 499~\AA\ and \sixii\ 521~\AA.

Although the UVCS synoptic program has changed with time, becoming more sporadic over the years, the spectra taken during the period under investigation always included the \ovi\ doublet, the two above-mentioned \sixii\ lines, and the \hi.
For better statistics, UVCS synoptic data were integrated with complementary special observations whenever the pointing was the same as the synoptic program. 
On average, the cadence of the data was about one per day, although unevenly spaced in time, with only a small number of gaps due to telemetry problems or special spacecraft maneuvers (for more details, see \citealt{Giordano2008}).
We used the latest version of the UVCS Data Analysis Software (DAS v. 51, available at the UVCS website),\footnote{{ https://www.cfa.harvard.edu/uvcs/get$\_$involved/das51.html}} for wavelength, intensity calibration, and removal of image distortion, which also takes into account the time variations of the instrument performances. 
The uncertainties in the line intensities are estimated to be about 20\% and are due to photon-counting statistics, background subtraction, and radiometric calibration (\citealt{Gardner2002}).
In order to estimate the total intensity of a selected spectral line, the calibrated and combined UVCS spectra were fit with a function resulting from the convolution of a Gaussian function (for the coronal spectral profile), with a Voigt curve describing the instrumental broadening and a function accounting for the width of the slit (\citealt{Giordano1998t}). 
While the two \sixii\ lines form by collisional excitation followed by radiative de-excitation, as usual for coronal lines, the \hi\ and \ovi\ coronal lines also form by resonant scattering of line radiation from the disk.
The temperature of maximum ion formation from the ionization equilibrium of the \hi\ line is about $\log_{10} (T/K) \approx 4.5$.
The \ovi\ lines peak at $\log_{10} (T/K) \approx 5.3-5.4$, while the \sixii\ lines peak at $\log_{10} (T/K) \approx 6.2-6.3$.
These five lines thus cover a wide range of temperatures in the corona.
Moreover, as already discussed, while \hi\ line intensities peak at streamer centers (especially along the equatorial streamer belt), the line intensities of \ovi\ peak at the edges of the streamers. 
\sixii\ lines add emission from high-density and high-temperature plasma associated with more localized active regions at higher latitudes, roughly corresponding to the location of sunspots on the surface of the Sun. 
In this work, we used data taken at a heliocentric height of 1.7~\rsun, due to the optimal coverage of the complete dataset at this height.
The instrument rolls analyzed in this work are the PAs corresponding to mid-latitude (45\dg, 135\dg, 225\dg, and 315\dg) and equatorial (90\dg\ and 270\dg) regions.
We remark that the data taken away from the center of the slit actually correspond to higher heliocentric distances (up to a few tenths of a solar radius).
To overcome this problem, the data at different angles from the center of the slit were obtained by interpolating the spectra taken with the slit positioned at different heights, to avoid possible effects on the radial dependence of the coronal rotation rate.
Finally, we only considered data in a range between 40\dg\ and 140\dg\ colatitude in order to avoid problems related to projection effects at higher latitudes (see discussion in \citealt{Mancuso2013}).

\section{Methodology}

The first step in our investigation relies on the identification of statistically significant periodic signals as a function of heliographic latitude.
In this section, we thus discuss the two main algorithms used in this work for the analysis of the whole dataset.

\subsection{Generalized Lomb-Scargle periodogram}

The Lomb-Scargle periodogram (LS; \citealt{lomb1976,Scargle1982}) is a technique that is widely used in astronomy to look for periodicities in datasets.
This was the main method used, along with the autocorrelation function, for detecting periodicities in the seminal work of \cite{Giordano2008}.
Adapted from the Fourier transform power spectrum, the LS periodogram excels at identifying periodic behavior in the frequency domain for data taken with uneven time sampling. 
Mathematically, the LS periodogram is based on searching for sinusoidal periodicities and calculates the goodness of fit of a sinusoidal function, as compared to the data, for a selected frequency grid.
Since the LS method assumes a zero-mean-harmonic model with constant noise variance, the data are to be pre-centered around the mean value of the observed values before performing the periodogram analysis. 
However, if the empirical mean (obtained by fitting the sample mean to the data) and the true mean differ significantly, this procedure can lead to incorrect period estimates.
For example, this assumption easily breaks down in the case of a small dataset or whenever the data does not uniformly sample all the phases, thus possibly incurring aliasing problems.
Several periodogram generalizations that are invariant to the shifts in the observed value have thus been proposed in the literature to address this specific problem.
In particular, a substantial improvement of the classical LS periodogram is contained in the work of \cite{Zechmeister2009} who extended the LS formalism to include weights for the measurement errors and constant offsets for the data.
This more refined procedure has been named the generalized Lomb-Scargle periodogram (GLS).
Compared to the classical LS periodogram, the GLS is known to provide a more accurate frequency determination, being less susceptible to aliasing and giving a much better determination of the spectral intensity.
The method takes into consideration measurement errors by introducing a weighted sum in the original LS formulation. 
Additionally, the GLS introduces an offset constant to overcome the assumption of the mean of the data.
In this technique, the harmonic model of the LS periodogram is directly extended with the addition of a constant offset as follows: 
\begin{equation}
    y(t) = \mathrm{A}\sin(\omega t) + \mathrm{B}\cos(\omega t) + \mathrm{C},
\end{equation}
where A, B, and C are constants calculated from the data, and the frequency, $\omega,$ is obtained by minimizing the squared difference between the observed data and the model function.
It has been shown that the GLS periodogram is more reliable than the classical LS periodogram in detecting periodicities when the light curve is not well sampled and the data sampling overestimates the mean (\citealt{Zechmeister2009,VanderPlas2015,VanderPlas2018}).

\subsection{Multichannel singular spectrum analysis}

The time series of the UV intensities are not strictly harmonic, meaning that a Fourier-based approach, such as the one adopted by the GLS algorithm, could be misleading in identifying the real periodicities in the datasets.
Differently from classical spectral analysis, where the basis functions are prescribed sinusoidal functions, a singular spectrum analysis (SSA; \citealt{Broomhead1986,Vautard1989}) produces data-adaptive filters that are capable of isolating oscillation spells, thus making this method more flexible and better suited for the analysis of nonlinear, anharmonic oscillations.
In particular, SSA can extract information and allow the identification of pure oscillatory signals from short and noisy time series without prior knowledge of the dynamics affecting the time series.
The original time-evolving signal is not simply decomposed into periodic sinusoidal functions as in Fourier-like techniques, but broken down into data-adaptive waves that can be modulated both in amplitude and phase. 
Although not commonly used in astrophysical contexts, this powerful technique has recently been applied to successfully detect Doppler-shift oscillations in the UV corona (\citealt{Mancuso2015,Mancuso2016}), analyze quasi-biennial oscillations of the \fe\ green coronal emission line at 5303 \AA\ (\citealt{Mancuso2018}), and, more recently, to detect multiple quasi-periodic pulsations observed during the flaring activity of a young, active solar-type star observed by the Kepler mission (\citealt{Mancuso2020}).
 
Its multivariate extension, the so-called multichannel singular spectrum analysis (MSSA), has the additional ability to identify coherent space-time patterns and thus extract common periodic signals, trends and noise from a multivariate dataset (e.g., \citealt{Taricco2015,Mancuso2018}). 
Moreover, an MSSA is much more flexible than the standard methods of modeling that involve at least one of the restrictive assumptions of linearity, normality, and stationarity.
The theoretical framework of MSSA was proposed initially by \cite{Broomhead1986} and later developed by \cite{Plaut1994} and \cite{Ghil2002}.
This data-adaptive technique includes two stages: decomposition and reconstruction. 
In the first stage, MSSA decomposes a multivariate time series $X_l(t)$, with $t=1, \ldots, N$ representing time and $l=1, \ldots, L$ the individual time series (or channels), into an orthonormal, data-adaptive space-time structure whose elements represent eigenvectors of a grand lag-covariance matrix of size $LM \times LM$, where $M$ is the width of a sliding $M$-point window. 
Diagonalizing the above matrix results in a set of $LM$ eigenvectors $\bm {E}^k$, with $1 \leq k \leq LM$, called space-time empirical orthogonal functions (ST-EOFs). 
Their associated space-time principal components (ST-PCs) $\bm {A}^k$, of time length $N' = N - M + 1$, are single-channel time series that are computed as
\begin{equation}
 A^k(t)  = \displaystyle\sum_{j=1}^{M}\sum_{l=1}^{L}  X_l(t+j-1) E^k_l(j),
\end{equation}
where $t$ varies from 1 to $N'$.
The $LM$ real eigenvalues $\lambda_k$, each associated with the $k$-th eigenvector $\bm {E}^k$, equal the variance in the $\bm {A}^k$ direction.
The ST-PCs thus represent the different oscillatory modes extracted from the dataset, although, because of the lag window, they cannot be located into the same index space with the original time series. 
It is possible, however, to represent the same information in the original coordinate system by means of the so-called reconstructed components (RCs). 
Thus, the $k$-th RC at time $t$ for channel $l$ is given by:
\begin{equation}
\scalebox{0.77}{
{\[R_l^k(t) =
\begin{cases}
\displaystyle {1 \over t}     \sum_{j=1}^{t}     A^k(t-j+1) E^k_l(j)  & \quad  (1 \leq t \leq M-1)   \\
\displaystyle {1 \over M}     \sum_{j=1}^{M}     A^k(t-j+1) E^k_l(j)  & \quad  (M \leq t \leq N-M+1) \\
\displaystyle {1 \over N-t+1} \sum_{j=t-N+M}^{M} A^k(t-j+1) E^k_l(j)  & \quad  (N-M+2 \leq t \leq N). \\
\end{cases}
\]}}
\end{equation}
Each RC allows the reconstruction of the dynamical behavior in $\bm {X}$ that belongs to $\bm {E}^k$.
By summing up all the individual RCs, it is finally possible to recover the original time series, so that information is not lost in the decomposition and reconstruction process.

An important characteristic of the MSSA technique is that it may be used to identify commonly modulated oscillations in the presence of colored noise (e.g., \citealt{Taricco2015,Mancuso2018}). 
In the single-channel case, any oscillation detected through a window of width $M$ can be completely described in terms of only two vectors, sine and cosine, with periods equal to the oscillation, provided that the period is less than $M$ and that the timescales of amplitude and phase modulation are much greater than $M$ (\citealt{Vautard1989}). 
If the variance of a series is dominated by such an oscillation, SSA will generate a pair of sinusoidal EOFs resulting in quadrature, that is, $\pi/2$ out of phase with each other and with similar amplitudes. 
Similarly, in the multichannel context, an oscillatory mode can be represented by MSSA as a pair of spatio-temporal patterns that are sinusoidal in time, have similar amplitudes, and are $\pi/2$ out of phase (\citealt{Plaut1994}).
After MSSA, a test of statistical significance is needed to avoid spurious oscillations that may be due to non-oscillatory processes, such as first-order autoregressive AR(1) noise.
The oscillatory modes identified with MSSA can be tested against a red noise null hypothesis through a Monte Carlo simulation (\citealt{Allen1996}).
This null hypothesis is that the data have been generated by $L$ first-order, autoregressive independent processes (i.e., red noise).
The dataset generated by the red noise model is called the surrogate dataset, and it is subjected to MSSA in the same way as the original dataset.
A large number of surrogates are generated to estimate the confidence limits for the MSSA result of the original dataset. 
The null hypothesis can be rejected if the spectrum of the eigenvalues associated with the modes detected by MSSA is higher than that expected in the data generated by red noise processes. 
As suggested by \cite{Groth2011}, we also relied on a subsequent Varimax rotation of the ST-EOFs to improve the separability of distinct frequencies.

\begin{figure*}
\centering
\includegraphics[width=12cm]{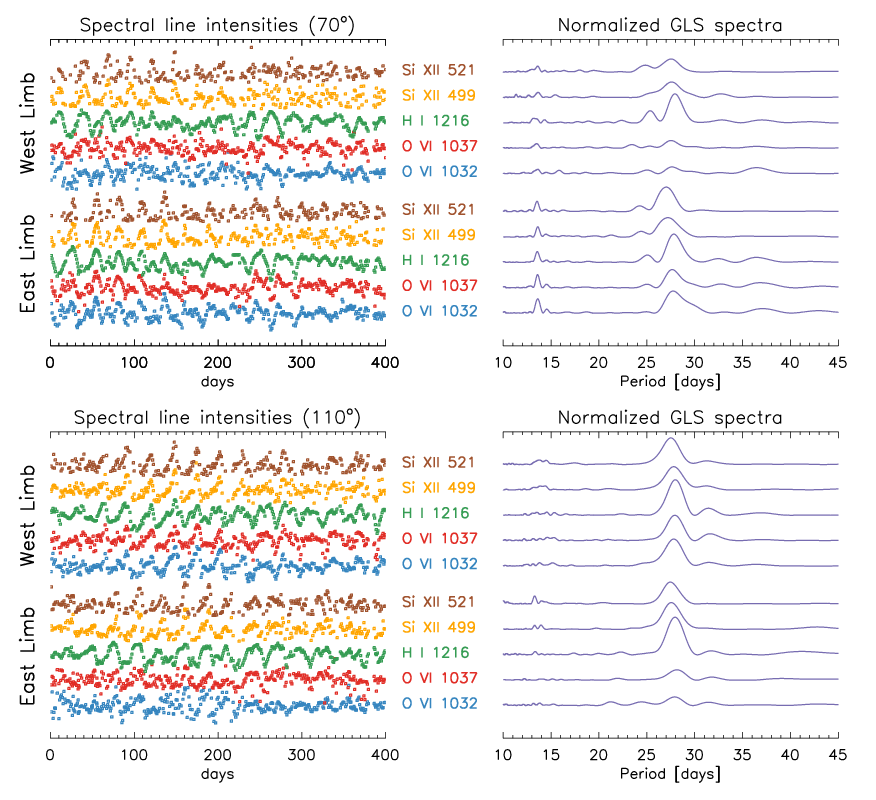}
\caption{Detrended time series of UV spectral line intensities used in this work at a colatitude of 70\dg\ over the 400-day interval from 1996 April 4 to 1997 May 15 ({\it top-left panel}) together with their respective normalized GLS spectra ({\it top-right panel}).
{\it Bottom panel}: Same as above but at a colatitude of 110\dg.}
\label{Fig2}
\end{figure*}

\section{Data analysis and results}

In what follows, we present the data analysis and results obtained by separately applying the two different techniques presented in the previous section.

\subsection{GLS periodogram}

We first performed the GLS periodogram analysis of the time-series data (for each UV spectral line, and, separately, for the east and west limbs) at steps of 5\dg\ from 40\dg\ to 140\dg\ in colatitude.
Trends were eliminated before the periodogram estimation via removal of best-fit second-order polynomials from the original time series, and by carrying out the GLS periodogram analysis on the detrended data.
Figure 2 shows examples of normalized GLS spectra obtained for the respective detrended intensity time series at colatitudes of 70\dg\ and 110\dg\ for both the east and west limbs.
The colatitude dependence of the rotation rate of the UV corona, obtained from the application of the GLS analysis at steps of 5\dg\ from 40\dg\ to 140\dg\ in colatitude, is presented in Fig. 3.
Uncertainties in the peak positions of the power spectra at each latitude were estimated using a Monte Carlo method over a large number of trials: an approximately Gaussian distribution of peak positions was obtained, and the uncertainty was estimated as its standard deviation (see \citealt{Giordano2008} for details).
One major challenge when interpreting periodograms arises when the noise in the time-series data produces various peaks in the resulting periodogram, some of which could be spurious. 
In this case, it is customary to use the false alarm probability (FAP) to assess the statistical significance of the highest peak in the periodogram.
The FAP is used to quantify the likelihood of a false detection of a signal and describes the probability that a peak in the power spectrum with a height above a certain level would occur purely by chance in the case of random white noise and the absence of that signal.
Assuming Gaussian noise, the FAP can be computed analytically by the following equation (\citealt{Horne1986}):
\begin{equation}
{\rm FAP} = 1 - (1 - e^{-z})^{N_i}, 
\end{equation}
where $z$ is the height of the corresponding peak in the periodogram, and $N_i$ is the number of independent frequencies in the time series.
The empirical formula above is adequate for our case because there is no significant level of data clumping (i.e., we don't have multiple observations per day). 
We ran the FAP tests on all periodograms and included in our analysis only those periods corresponding to peaks in the periodograms that were considered significant.
As a general rule, a FAP value $< 0.1$\% was chosen to indicate that the peak was likely to be significant.

\begin{figure*}
\centering
\includegraphics[width=16cm]{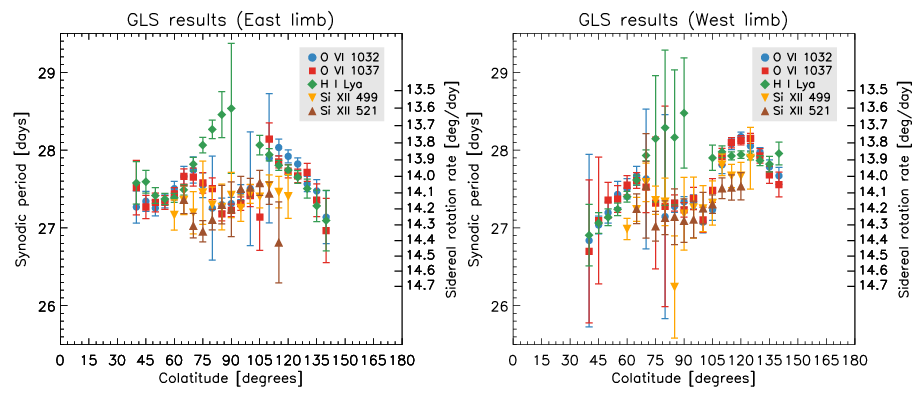}
\caption{Colatitude dependence of UV corona rotation rate at 1.7 \rsun\ obtained with the generalized Lomb-Scargle (GLS) periodogram technique at the east ({\it left panel}) and west limbs ({\it right panel}) from the analysis of the time series of five different UV spectral lines. }
\label{Fig3}
\end{figure*}

One first important aspect of the periodogram analysis on the single intensity time series as evinced from Fig. 2 is that the normalized GLS spectra do vary depending on both the (east or west) limb and the (north or south) hemisphere locations.
For example, the \ovi\ doublet spectra at 70\dg\ of colatitude in the east limb are much more significant than the ones at the west limb; vice versa, the \ovi\ doublet spectra at 110\dg\ of colatitude in the west limb are much more significant than the ones at the east limb.
Moreover, from visual inspection of Fig. 3, we can immediately appreciate the different colatitude dependence of the rotation rate of the UV corona in the east and west limbs, as evinced by applying the GLS technique to the intensity time series from the five different UV spectral lines. 
This difference doesn't appear to be dependent on the specific spectral line under consideration, within the given uncertainties, which could hint more to a viewing or systematic instrumental effect rather than to stochastic effects that may be related to poor statistics. 
Notwithstanding the above, this is an unexpected outcome.
North-south rotational asymmetry has actually been physically interpreted in the framework of a model of interdependence of solar rotation and activity (\citealt{Brun2004,Brajsa2006}) due to the oft-reported asymmetric latitudinal distribution of solar indices of activity, and has been in fact reported by several authors in the literature (e.g., \citealt{Hansen1969,Hoeksema1987,Obridko2001,Zaatri2009,Sykora2010,Vats2011,Gigolashvili2013}). 
However, east-west asymmetry has been seldom reported also because, while it is certainly difficult to justify and interpret from a physical point of view, it does foster suspicions on the reliability of the data at hand.
This apparent east-west asymmetry, already noticed by \cite{Giordano2008} and also mentioned by \cite{Lewis1999} in their analysis of white-light data related to the same solar minimum period, can be tentatively attributed to a viewing effect arising from the inclination of the magnetic field axis with respect to the rotation one (e.g., \citealt{Mancuso2007}) mixed with projection effects at higher latitudes (\citealt{Mancuso2013}).
As a matter of fact, this is a major unsolved problem in the interpretation of our results as evinced from the GLS periodogram analysis of the singular intensity time series, and it does cast legitimate doubts on the robustness of the results obtained by \cite{Giordano2008}. 
It would be reasonable to expect similar latitudinal profiles at both east and west limbs, although the two sets of profiles do appear to marginally agree within the given uncertainties. 
In view of the fact that it is daring to argue that the observed east-west asymmetry represents a physical phenomenon, this puzzling result may hint to major sources of errors in the separate analyses of the two limbs.
A different approach to the analysis of the coronal rotation, such as one that would take into account the expected anticorrelation between the intensity variations simultaneously observed in the two limbs and the multivariate analysis of different spectral lines, is thus to be looked for in order to eliminate this bias. 
Such a novel method is presented, applied, and discussed in next section.

Another puzzling aspect is that all the rotation rates derived from the analysis of the UV spectral lines considered in this work do appear to agree fairly well in their values, within the given uncertainties, apart from the rotation rate of the \hi\ line, which is definitely much smaller at lower latitudes (say, within 20\dg\ from the equator), especially in the northern hemisphere.
This effect appears to be real and it is especially noticeable in the east limb, whereas the uncertainties in the rotation rate of the \hi\ line on the west limb are probably too large to allow us to draw any firm conclusion.
A partial explanation for this apparent inconsistency can be attributed to the different dependence of the intensity of the individual spectral lines and the fact that the intensities are obtained by integrating contributions all along the line of sight throughout the extended corona.
UV lines in the lower corona form essentially by collisional excitation followed by radiative de-excitation, but they also contain a resonance scattering component that is more or less important depending on the characteristics of the spectral line. 
At greater heights, and for bright enough exciting chromospheric radiation, resonant scattering is the dominant process for the hydrogen \hi, while both components need to be considered for the \ovi\ lines (the \sixii\ lines are almost completely collisional).
The important point is that while the collisional component depends on electron densities, $n_e$, via the line-of-sight integral of $n_e^2$, the radiative component scales as $n_e$.
This translates to the fact that the emission from spectral lines where the collisional component is predominant is deemed to be more concentrated on the plane of the sky (where distances from the Sun's surface are lower) and necessarily reflect rotation periods of high-density coronal structures.
Vice versa, emission from spectral lines where the radiative component is predominant, such as the \hi\ line, includes contributions from plasma distributed over a larger integration path and is therefore expected to be less influenced by high-concentration plasma structures associated with active regions, thus reflecting the quieter plasma. 
The different dependence of the line intensities on the line-of-sight integrals of $n_e$ can thus be connected, in a currently unascertained way, to the above apparent inconsistency between the behavior of the \hi\ line and the other UV spectral lines within the equatorial streamer belt. 
A less troublesome explanation for the striking difference observed at equatorial latitudes in the rotation rate of the \hi\ emission (as compared to the other UV spectral lines) may lie in the fact that strong dimming in the core of the streamer belt is detected only in the case of the \ovi\ and \sixii\ lines. 
This reveals a local depletion of oxygen and silicon ions with respect to the photosphere (e.g., \citealt{uzzo2003}). 
As suggested by \citet{noci1997b}, this effect could be due to the geometry of the flux tubes of open magnetic field lines separating the multiple loop structures inside the core of complex streamers. 
More specifically, the narrowing of the cross-section of the flux tubes guiding outflows of coronal plasma from an initially high areal divergence has the effect of slowing down the solar wind relative to a radial flow. 
A wind speed decrease with unchanged density implies a reduced proton flux. 
Since oxygen and silicon ions are dragged into the solar wind via Coulomb collisions by a force proportional to the proton flux (\citealt{geiss1970}), a reduction in this flux causes a decrease in the oxygen and silicon abundance, which is observed as an \ovi\ and \sixii\ dimming in the streamer core, but not in the hydrogen abundance.
In fact, the above mentioned effect is probably due a concurrence of causes, such as the higher diffusion of the light H atoms, the different relative abundances of the H and O atoms in the streamer belt, and the different sensitivities of the \ovi\ and \hi\ spectral lines to the electron density along the line of sight (\citealt{Noci1997}). 
In any case, both the unexpected different profiles of differential rotation for the two limbs and the discordance between the results from different lines cast legitimate doubts on the validity of periodogram approaches, such as LS or GLS, in analyzing singular spectral-line intensities for periodicity searches in the optically thin corona.

\subsection{MSSA}

A new, alternative approach to circumventing the problems expounded in the previous section is the use of an independent multivariate analysis that can extract common oscillatory modes and take into account the expected anticorrelation between the intensity time series of the two limbs. 
Previous studies have shown that MSSA, which was presented in an earlier section, is a data-adaptive technique that is particularly successful at reliably isolating even weak common oscillations in short and noisy time series.
In order to apply the MSSA algorithm, all intensity time series have been linearly interpolated in time to a regularly spaced array with a time interval of one day as the original time series.
Furthermore, to avoid the dominance of variability in one or several channels, each time series was centered to have zero mean and unit variance (\citealt{Ghil2002,Mancuso2018}).
No other filtering or detrending was applied to the data.
In MSSA, the length of the lag window $M$ is a user choice and is generally selected in view of a trade-off between spectral resolution and statistical significance of the obtained components.
Simply put, if $M$ is large, more temporal information can be extracted, but at the same time the variance is distributed on a larger set of components. 
On the contrary, if $M$ is small, the statistical significance of the obtained components is enhanced. 
In general, the stable features of the set of eigenvalues and eigenvectors can be evaluated by varying the window size $M$ over a given range, $M_1 \leq M \leq M_2$.
However, the results of MSSA do not change significantly with varying $M$ as long as $M \ll N$ (\citealt{elsner1996}).
The use of a lag window length $M$ typically allows the distinction of oscillations with periods in the range $[M/5, M]$ (\citealt{Plaut1994}).
Different window lengths can thus be used to allow verifying how much the decomposition is sensitive from the choice of this parameter, although greater statistical confidence is expected for smaller window lengths (\citealt{Ghil2002}).
Since the use of a given $M$ allows the identification of oscillations with periods that do not exceed $M$, we expect to  reliably identify $\sim 27$-day periodicities by choosing three different window lengths corresponding to $M=40,60,$ and 80 days.

\begin{figure}[t]
\centering
\includegraphics[width=8cm]{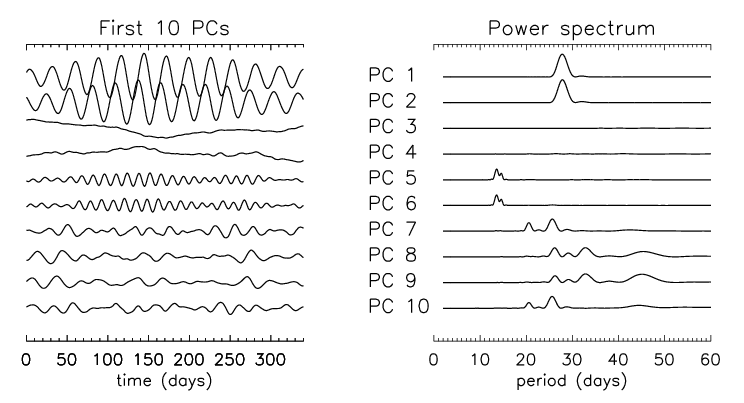}
\caption{Ten leading temporal ST-PCs at a colatitude of 120\dg\ ({\it left panel}) together with their Fourier power spectra ({\it right panel}). The lag window length $M$ used in MSSA is $M=60$ days. These PCs are time coefficients that weigh the corresponding T-EOFs in reconstructing the original time series; they represent the different modes of variability of the time series.}
\label{Fig4}
\end{figure}

As an example, Fig. 4 (left panel) shows the ten leading ST-PCs obtained by applying MSSA to the ten series at a colatitude of 120\dg\ with a lag window length $M=60$ days, together with Fourier power spectra of the respective ST-PCs (right panel in Fig. 4).
Following the methodology presented above, the statistical significance of the retrieved oscillations (whose ST-PCs resulted in quadrature, that is, $\pi/2$ out of phase with each other and with similar amplitude) was tested by Monte Carlo methods in relation to the null hypothesis of red noise at the 5\% significance level (one-tailed test). 
Results are presented in Fig. 5 for the case at colatitude 120\dg.
The resulting singular spectrum, shown in the inset of the same figure, displays the eigenvalues sorted in decreasing order and normalized so that they represent their corresponding portion of the total variance of the time series. 
Most of the variance in the time series is contained in the first two eigenvalues, with a sudden drop and a plateau after the fourth one. 
The four leading eigenvalues represent both the 27-day signal (ST-PCs 1-2) and trend (ST-PCs 3-4) contained in the time series, while the higher ranked ones correspond to noise. 
We note that, differently to previous methods used while reducing the data with the GLS technique, the trend was not removed from the original datasets by imposing a low-order polynomial fit to the data, but it was naturally found as a pair of leading PCs (ST-PCs 3-4) because of the data-adaptive nature of MSSA.
The first two ST-PCs (ST-PCs 1-2) that capture the $\sim 27$-day signal are nearly periodic, with the same period, and, as expected by theory, in quadrature with each other (see Fig. 4).
We remark that for all colatitude angles between 40\dg\ and 140\dg, the $\sim 27$-day oscillation, corresponding to the first two ST-PCs, was statistically significant at the chosen significance level (see Fig. 5). 
As an illustration, in Fig. 6 (top panel) we show a contour plot of the original UV line intensities versus time ($x$-axis) and spectral line ($y$-axis) at a given colatitude 
(120\dg) for both the east and west limbs. 
In the bottom panel of Fig. 6, we show a similar contour plot, but it is derived from the UV line intensities reconstructed from the first two leading components (RCs 1-2) obtained with the MSSA technique by using a lag window length $M=60$.
From visual inspection, it is evident that the $\sim 27$-day variability of the UV intensity in the latter plot is clearly enhanced for each spectral line, thus allowing a more accurate determination of the involved periodicities. 

\begin{figure}
\centering
\includegraphics[width=8cm]{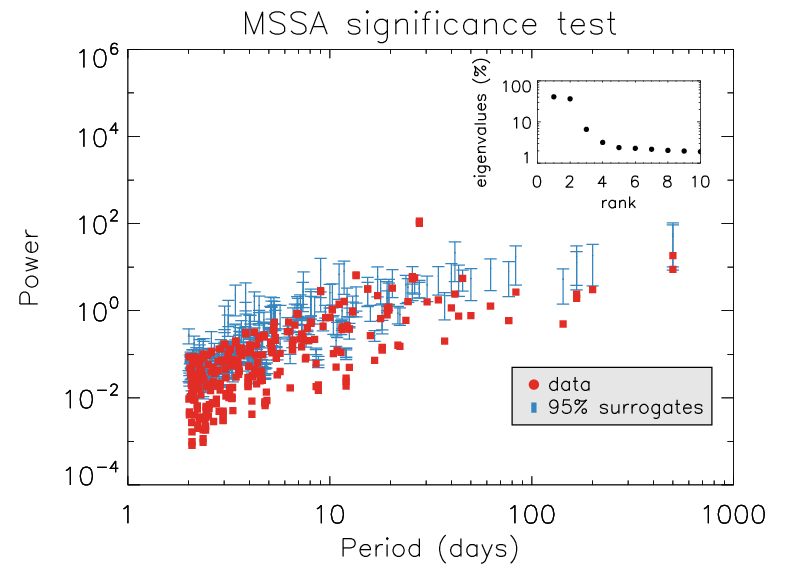}
\caption{Monte Carlo MSSA test of the ten time series at a colatitude of 120\dg.
The lag window length $M$ is 60 days.
The significant signals (at 5\% significance level) are those whose data eigenvalues lie above the 97.5th percentiles of the surrogate eigenvalues: according to the test, these signals have more variance than would be expected from a noise process.
Eigenvalues (shown in the inset) are normalized to represent the appropriate fraction in \% of the total variance of the time series.}
\label{Fig5}
\end{figure}

\begin{figure}
\centering
\includegraphics[width=8cm]{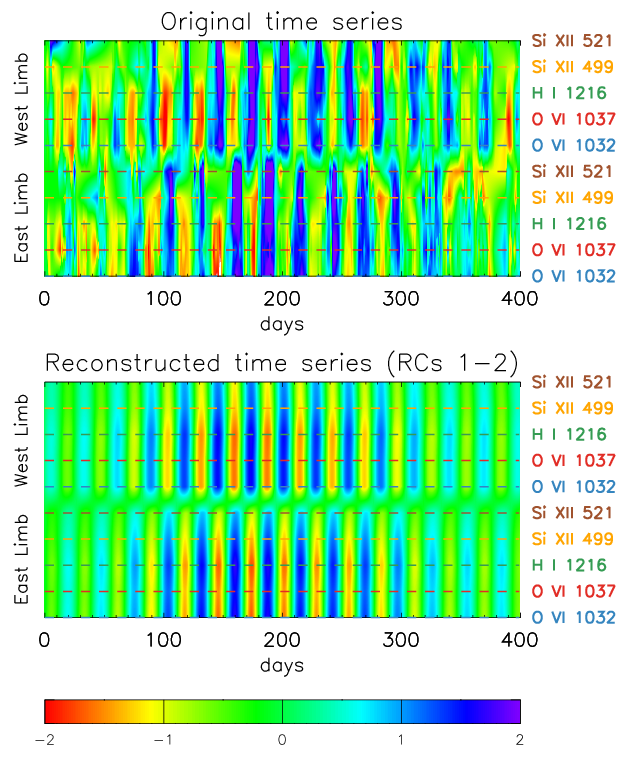}
\caption{
{\it Top:} contour plot showing original UV line intensities versus time ($x$-axis) and spectral line ($y$-axis) at a given colatitude (120\dg) for both east and west limbs.
{\it Bottom:} same as above but showing the UV line intensities reconstructed from the first two leading components (RCs 1-2) obtained with the MSSA technique by using a lag window length $M=60$.
}
\label{Fig6}
\end{figure}

\begin{figure*}
\centering
\includegraphics[width=12cm]{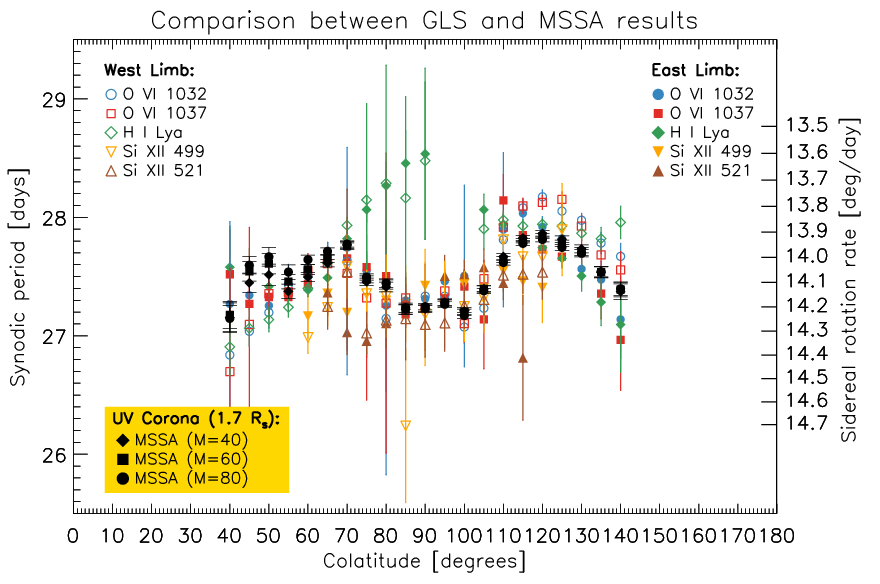}
\caption{Colatitude dependence of coronal rotation rate at 1.7 \rsun\ obtained with the Lomb-Scargle periodogram technique from the analysis of the time series from five different UV spectral lines and from the MSSA technique with three different windows of lengths $M = 40,60,80$.}
\label{Fig7}
\end{figure*}

\section{Discussion}

A comparison of the results obtained from the application of the two distinct methods (GLS and MSSA) to the complete set of data is displayed in Fig. 7.
In this plot, showing the colatitude dependence of the coronal rotation rate at 1.7 \rsun, we superimposed all results obtained from the GLS periodogram analysis of the intensity time series (from both the east and west limb), together with the result obtained from the MSSA analysis of the spectra from both limbs applied with the three selected window lengths $M$.
Although the MSSA result is almost independent of the choice of $M$, we included all three cases to better quantify how the choice of the window length, being a free parameter of the algorithm, influences the uncertainty of the MSSA (at most 0.1\dg\ day$^{-1}$, but more in the northern than in the southern hemisphere).
Interestingly, the general profile of the rotation rate as a function of colatitude appears to be less asymmetric than evinced by \cite{Giordano2008} and by the GLS periodogram applied to the full dataset (see also Fig. 3). 
We discuss the implications of this outcome further after comparing our results with other concomitant observations taken around the same period. 

Comparisons of differential rotation profiles of the corona are often presented in the literature by taking profiles obtained at different times during the same solar cycle (or even different ones) and various techniques as references.
However, these profiles are known to vary along the solar cycle due to the activity of the Sun (e.g., \citealt{Mancuso2011}) and probably even along several solar cycles (\citealt{Brajsa2006}).
To allow for a proper comparison with other datasets concerning the latitudinal dependence of the rotation rate below and above the solar surface, our discussion is limited to studies utilizing data obtained in a time interval roughly corresponding to the same period considered in this study. 

Magnetic activity on the Sun is not uniformly distributed over the solar surface, but is instead concentrated into active regions of which sunspots represent an essential component.
In the case of the observation obtained with the UVCS instrument, the rotational signal derived from the analysis of the intensity time series of different UV spectral lines can be mostly attributed to the rotation of individual large-scale coronal structures, such as streamers, mostly linked to active regions passing across the solar limbs. 
It is thus reasonable to assume that the rotation rate measured in the UV corona reflects the rotation of the large sunspot groups over which large active region streamers are known to develop. 
This is especially true in the case of the hotter \sixii\ lines, whose signal is generally enhanced at the mid-latitudes corresponding to the surface location of sunspots.
Sunspots and sunspot groups have been used for centuries as tracers for solar rotation, so they represent one of those most commonly used in the literature due to the availability of long-term datasets from various observatories.
\cite{Brajsa2006}, using rotation-rate residuals calculated from sunspot groups, found a secular deceleration of rotation and faster rotation at the minimum than at the maximum of the solar cycle, thus implying that a proper comparison should be made with data taken during the pertinent solar minimum.
Unfortunately, during a solar minimum, sunspot data are unavoidably scarce, thus leading to large statistical errors.
\cite{Ruzdjak2017} examined the sunspot position from various databases in the 1874 to 2016 period to calculate yearly values of the solar differential rotation parameters, A and B, and reported values and errors of differential rotation coefficients for several solar cycle minimum and maximum years covered by the data.
We remind the reader that the latitudinal variation of the rotation rate in the solar context has traditionally been represented by a three-term expression:
\begin{equation}
\Omega(\theta) = \mathrm{A} + \mathrm{B}\cos^2{\theta} + \mathrm{C}\cos^4{\theta}, 
\end{equation}
where $\Omega$ is the angular velocity, $\theta$ is the solar colatitude, and A; B; and C are parameters obtained by fitting this function to the measured rotation rates at each latitude.
In this formula, coefficient A represents the solar rotation rate at the equator, whereas coefficients B and C account for the departure of rigid rotation as a function of latitude.
\cite{Ruzdjak2017} reported sidereal values (in [\dg\ day$^{-1}$]) of $A = 14.70 \pm 0.09$ and $B = -2.6 \pm 1.2$ for data taken during the minimum of the year 1996 from the Debrecen Photoheliographic Data (DPD) database (\citealt{Sudar2017}).
In Fig. 8, we compare this profile (blue dotted line and shaded area) with the one obtained in this work by means of the data-adaptive MSSA technique (black points with error bars). 
Although the curve retrieved from the sunspot group data has large uncertainties, it is obvious that sunspots do appear to rotate much faster near the equator than the coronal structures sampled by UVCS, at least within 20\dg\ above and below the solar equator. 
On the other hand, as for the profile beyond this interval, the two sets of data are marginally compatible, although no hint of a deceleration of the rotation with higher latitudes emerges from the UV data that clearly manifest a quasi-rigid body rotation. This is in agreement with previous investigations on the differential rotation of the extended corona (\citealt{Timothy1975,Bohlin1977,Hansen1969, Parker1982,Fisher1984,Weber1999,Lewis1999}).

\begin{figure*}
\centering
\includegraphics[width=12cm]{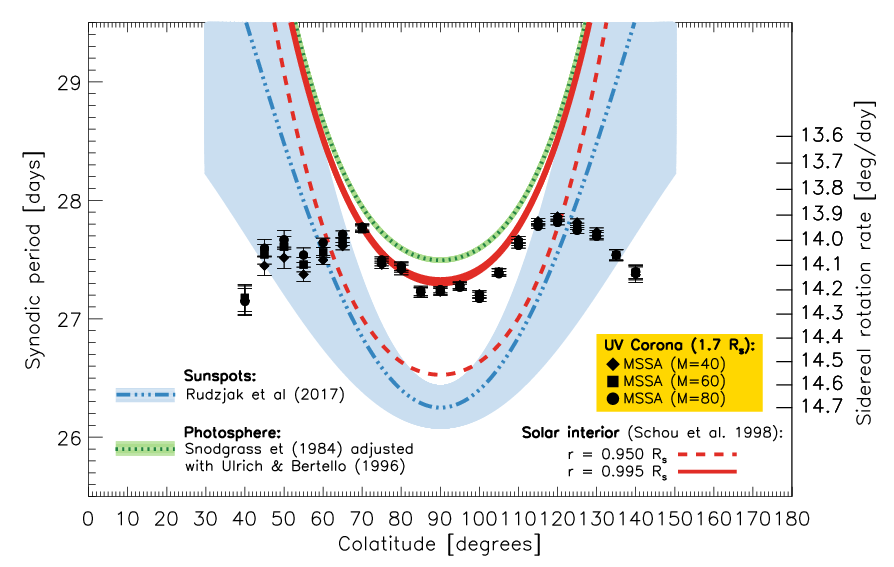}
\caption{Comparison of results obtained for the differential rotation of the V corona in this work with the MSSA technique and from other authors.}
\label{Fig8}
\end{figure*}

\cite{Snodgrass1984} determined the differential rotation of the photospheric plasma from Doppler shifts of the {Fe~{\sc{i}}} spectral line at 5250 \AA\ obtained at the Mt. Wilson Observatory (MWO) during the 1967 -- 1984 period.
This differential profile has definitely become a standard in the literature for comparing the solar rotation rates obtained through different techniques and at different heights from the solar surface.
However, as in the case of sunspots, several authors seem to indicate persistent modulations of the rotation rate during the solar cycle, so that the above rate has to be adjusted depending on the phase of activity of the Sun.
\cite{ulrich1996} reported high precision solar rotation rate measurements obtained with the MWO telescope system up to the end of 1995, when the activity of the Sun was already near its minimum, yielding a sidereal equatorial rotation rate value of 2.844 $\mu$rad s$^{-1}$ corresponding to A = 14.079\dg\ day$^{-1}$.
The comparison of the \cite{Snodgrass1984} -adjusted with the \cite{ulrich1996} equatorial rotation rate value; green dotted line in Fig. 8- with our results clearly indicates that the UV corona rotates faster at all latitudes than the photospheric plasma, especially above about $\pm 20$\dg\ from the equator.
We thus conclude that the UV corona rotates more rigidly than the photosphere, with the corona rotating faster than the photosphere at all heliographic latitudes. 

The rotation rate of large, long-lived sunspots is certainly related to a large-scale rotation in that part of the convection zone, in the interior of the Sun, where these magnetic structures are anchored.
Vice versa, rotation rates of smaller, short-lived magnetic structures cannot be considered as linked to the convective region as a whole, since they are heavily affected by local and transient motions in the ambient solar active regions.
According to the previous discussion, since streamers develop over active region groups, we would expect their rotation curves to be correlated. 
However, if the rotation rate of the UV corona reflects the anchoring of the streamers responsible for the observed spectral line emission, it is evident that the roots of these coronal structures differ from the roots of long-lived sunspots.
This issue can be effectively addressed by comparing the above profiles with the ones obtained during a similar time interval via helioseismology studies.

The Sun is known to support a rich spectrum of internal waves continuously excited by turbulent convection.
Helioseismology provides a set of tools for probing the solar interior in three dimensions using measurements of wave travel times and local mode frequencies.
A major achievement of global helioseismology is the inference of the angular velocity in the solar interior through which it became possible to determine the rotation profile of the solar interior as a function of radius and latitude. 
These studies have shown that the Sun exhibits a complex differential rotation profile, with a differentially rotating convection zone and a rigidly rotating radiative interior separated by a transition region at about 0.7 \rsun, the tachocline, which is supposed to be the seat of the solar dynamo (see \citealt{Howe2009} for a comprehensive review).
\cite{Schou1998} carried out a comprehensive analysis of the rotation profile of the solar interior over 144 days, spanning May to September of 1996, from measurements of $p$-mode frequency splittings via comparison of several different rotation inversion techniques.
For their work, they considered Doppler imaging observations provided by the Solar Oscillations Investigation (SOI) using the Michelson Doppler Imager ({\it SOHO}/MDI).
Their results implied that the rotation in the bulk of the convection zone above the tachocline had a slow increase with radius at most latitudes, with the maximum rotation rate, 470.1 nHz (14.6\dg\ day$^{-1}$), being found on the equator at $r \approx 0.94$ \rsun.
Interestingly, this result matches (within the quoted uncertainties) the rotation rate of the sunspots outlined by \cite{Ruzdjak2017} quite well, thus suggesting that the roots of large-scale sunspots might be anchored at about just that depth in the interior of the Sun. 
In fact, it is not unreasonable to suggest that the roots of very large coronal structures may be anchored deep below the surface (see discussion in \citealt{Hiremath2013}).

Above this depth ($r \approx 0.94$ \rsun), toward the surface, a thin shear boundary layer was\ notably ascertained in which the rotation rate is seen to rapidly decrease at intermediate and high latitudes.
Since the short-lived magnetic structures characterizing the photospheric plasma rotate with a slightly slower rotation rate than observed by SOI/MDI data at 0.995 \rsun\ (but much slower than the rotation rate at 0.95 \rsun), the coronal structures responsible for the UV emission observed by UVCS in the corona may be rooted just below the surface of the Sun (say, around 0.99 \rsun) at the top of the convection zone. 
This suggestion was already given by \cite{Mancuso2012}, who discovered a striking significant positive correlation ($r = 0.629$ at 0.99 \rsun) between the variations in the residual rotation rates of the coronal and subphotospheric equatorial plasma all along solar cycle 23.
In this work, we not only confirm the above discovery, but extend its validity to an equatorial region as wide as about 40\dg\ in latitude centered around the equator.
The different differential rotation rate of large-scale sunspots as compared to the UV corona is thus explained if we posit that the streamer structures responsible for the UV emission are intermediate-scale structures anchored to lower depths than the large-scale sunspots.

The results from the above analysis obtained from UVCS spectral lines also reveal a mixture of differential rotation within about $\pm20$\dg\ of the equator, coupled with a quasi-rigid rotational profile at higher solar latitudes.
Following \cite{Nash1988}, the outer coronal field is supposed to rotate more rigidly than the underlying photosphere because it only depends on the lowest order harmonic components.
Coronal holes are known to represent extended regions of the open magnetic field with densities significantly lower than the typical background corona that appear to rotate more like a solid body than differentially (\citealt{Timothy1975}).
During the solar maximum, coronal holes are distributed at all latitudes. 
At solar minimum, however, they mainly occur near the polar regions.
Since the high-latitude regions covered by coronal holes during solar minimum activity consist of low-order magnetic multipoles, the open flux necessarily rotates more rigidly than the bulk of the photospheric field dominated by high-order multipoles (\citealt{Wang2004}).
A possible explanation for this anomalous behavior is that the motion of the coronal hole boundary uncouples from the motion of the underlying photospheric field due to continual interchange reconnection between open field lines at the hole boundaries and neighboring loops (\citealt{Nash1988,Wang1989}).
In this way, coronal holes are supposed to maintain their quasi-rigid rotation by adjusting their boundaries via reconnection with the outside magnetic field, while the underlying photosphere in which their open field lines are rooted, continues to rotate differentially.
Supportive observational evidences for the presence of small-scale, low-energy reconnection events along CH boundaries were recently presented (e.g., \citealt{Madjarska2004,Ma2014,Kong2018}).
It is then possible that the larger-scale and longer lived fields are rotating more rigidly than the small-scale fields, and it is these larger scale fields (\citealt{Hoeksema1987,stix1989}) that most strongly influence the intermediate-scale coronal structures responsible for the UV emission above the equatorial streamer belt.

The double nature of the differential profile of the rotation rate of the UV corona can thus be explained within the framework of a model in which the rotation rate at mid and high latitudes is influenced by large-scale coronal structures linked to the rigidly rotating coronal holes, while the rotation of the corona in the equatorial belt is mostly influenced by the differential rotation of shorter lived, intermediate-scale magnetic bipole structures anchored in the near-surface convection zone.
We further mention that \cite{Hiremath2013} claimed that magnetic reconnection alone cannot sufficiently explain the rigid body rotation and area evolution of coronal holes, suggesting that they must be deep rooted rather than mere surface phenomena.
Since the rotational profile of the solar interior flattens just below the tachocline, reaching sidereal rotation rates around 13.8\dg\ day$^{-1}$ (e.g., \citealt{Beck2000}), this hypothesis would easily explain the observed rigid rotation of the UV corona (which is supposed to be anchored deeper and deeper as a function of latitude). 
However, there are currently no accepted models of magnetic field generation that could anchor coronal structures to such a depth in the interior and, though fascinating, this idea remains just an alluring conjecture.

A final aspect we would like to discuss is the apparent north-south asymmetry especially evinced from the results obtained at the west limb with the GLS technique (see Fig. 2).
As already discussed, north-south rotational asymmetry can be physically interpreted in the framework of a model of interdependence of solar rotation and activity (\citealt{Brun2004,Brajsa2006}). 
During the first part of the ascending phase of solar cycle 23, the southern solar hemisphere was more active (\citealt{Temmer2006}), and,  on average, the stronger magnetic activity could have suppressed the differential rotation in that hemisphere. 
This is due to Maxwell stresses, related to the magnetic field, which tend to oppose Reynolds stresses, which are the main cause of the differential character of the solar rotation.
Although the results obtained by applying the GLS periodogram technique to the intensity time series of the five different UV spectral lines matches the result obtained by \cite{Giordano2008}, the application of the more sophisticated MSSA technique used in this work actually reveals that the alleged north-south asymmetry is probably only marginally significant. Additionally, it could actually be stochastically enhanced by problems related to poor statistics due to the use of a single set of observations obtained by the analysis of a specific spectral line.
As already outlined, individual subtraction of the unknown trend for each time series by an arbitrary analytical function may actually affect the derived period, so a data-adaptive trend removal obtained by a multivariate approach should certainly be preferred.
In fact, we also tried to subtract different trend functions from the data and found that the period determination is often affected (albeit mildly) by the chosen order of the polynomial.
Finally, any controversial hypothesis linked to the east-west asymmetry observed in the GLS analysis is automatically solved with the MSSA technique that naturally takes into consideration the expected anticorrelation between the periodical signals of the two limbs.
\section{Summary and conclusions}

We analyzed the UV coronal emission at 1.7 \rsun\ during the solar minimum preceding solar cycle 23 to investigate the differential rotation rate of the solar corona by means of a set of 400-day-long UV spectral line intensities of five different spectral lines: \ovi\ 1032 \AA, \ovi\ 1037 \AA, \sixii\ 499 \AA, \sixii\ 521 \AA, and \hi\ 1216 \AA, routinely observed by the {\it SOHO}/UVCS instrument.
For this goal, we used two different techniques: the GLS periodogram and a data-adaptive, multivariate technique, MSSA. 
This work integrates and extends the previous analysis from \cite{Giordano2008} that used the LS periodogram and the autocorrelation technique to analyze a 365-day-long set of data taken from the \ovi\ 1032 \AA\ spectral line.
With respect to that work, we improved our analysis via several methods: 
i) by analyzing data from both limbs taken using five different UV spectral lines; ii) by extending the dataset from 365 days to 400 days; iii) by working on newly calibrated data; iv) by employing a data-adaptive technique that takes into consideration the expected anticorrelation between the data at the east and west limb; and v) by detrending the data without assuming a pre-ordered low-order polynomial. 

One major conclusion of our analysis is that the low-latitude region of the UV corona (below about $\pm 20$\dg\ from the solar equator) shows differential rotation, while the higher latitude structures rotate quasi-rigidly.
The profile of the rotation rate at low heliographic latitudes is compatible with the rotational profile inferred in the interior of the Sun in the near-surface convective zone, thus suggesting the emission of UV radiation in the corona is linked with intermediate-scale magnetic bipole structures anchored near 0.99 \rsun.
The quasi-rigid rotation rate found at mid and high latitudes is instead attributed to the influence of large-scale coronal structures linked to the rigidly rotating coronal holes.
With respect to the previous analysis of \cite{Giordano2008}, the new MSSA analysis shows that rotation periods in the northern hemisphere are not very different from those of the southern hemisphere. Thus, the north-south rotational asymmetry, if really existent, is probably less pronounced than previously inferred. 

Independently from the specific application to the UV corona, we believe that the approach presented in this paper could represent a milestone for future investigations on the differential rotation of the Sun when dealing with simultaneous observations of coronal spectral lines. 
In particular, we are planning a similar inquiry that will involve a multiwavelength study of the rotation of the inner corona as evinced by a data-adaptive, multivariate analysis of the intensity variations detected in all coronal extreme ultraviolet (EUV) channels of the Atmospheric Imaging Assembly (AIA; \citealt{Lemen2012}) on-board the {\it Solar Dynamic Observatory} ({\it SDO}, \citealt{Pesnell2012}). 
In conclusion, we have introduced an alternative approach to extracting and describing the differential profile of the coronal rotation rate from coronal UV spectral line intensity time series, thus showing that MSSA is a powerful and complementary tool for exploring the spatio-temporal behavior of the coronal emission.

\begin{acknowledgements} 
We thank the referee for the useful comments on the manuscript. 
SOHO is a project of international cooperation between the European Space Agency (ESA) and the NASA. UVCS is a joint project of NASA, Italian Space Agency (ASI) and the Swiss Funding Agencies.
UVCS data were obtained from the {\it SOHO} Archive at the website https://sohowww.nascom.nasa.gov/data/archive/index$\_$gsfc.html).
We are also grateful to the TCD group at UCLA for the use of the SSA-MTM Toolkit (https://dept.atmos.ucla.edu/tcd/ssa-mtm-toolkit).
\end{acknowledgements}

\bibliographystyle{aa}
\bibliography{Mancuso_et_al_2020_ArXiv}

\begin{thebibliography}{76}
\expandafter\ifx\csname natexlab\endcsname\relax\def\natexlab#1{#1}\fi

\bibitem[{{Abbo} {et~al.}(2019){Abbo}, {Giordano}, \& {Ofman}}]{Abbo2019}
{Abbo}, L., {Giordano}, S., \& {Ofman}, L. 2019, \aap, 623, A95

\bibitem[{{Allen} \& {Robertson}(1996)}]{Allen1996}
{Allen}, M.~R. \& {Robertson}, A.~W. 1996, Climate Dynamics, 12, 775

\bibitem[{{Beck}(2000)}]{Beck2000}
{Beck}, J.~G. 2000, \solphys, 191, 47

\bibitem[{{Bohlin}(1977)}]{Bohlin1977}
{Bohlin}, J.~D. 1977, \solphys, 51, 377

\bibitem[{{Braj{\v{s}}a} {et~al.}(2006){Braj{\v{s}}a}, {Ru{\v{z}}djak}, \&
  {W{\"o}hl}}]{Brajsa2006}
{Braj{\v{s}}a}, R., {Ru{\v{z}}djak}, D., \& {W{\"o}hl}, H. 2006, \solphys, 237,
  365

\bibitem[{{Braj{\v{s}}a} {et~al.}(2004){Braj{\v{s}}a}, {W{\"o}hl},
  {Vr{\v{s}}nak}, {Ru{\v{z}}djak}, {Clette}, {Hochedez}, \&
  {Ro{\v{s}}a}}]{Brajsa2004}
{Braj{\v{s}}a}, R., {W{\"o}hl}, H., {Vr{\v{s}}nak}, B., {et~al.} 2004, \aap,
  414, 707

\bibitem[{{Broomhead} \& {King}(1986)}]{Broomhead1986}
{Broomhead}, D.~S. \& {King}, G.~P. 1986, Physica D Nonlinear Phenomena, 20,
  217

\bibitem[{{Brun} {et~al.}(2004){Brun}, {Miesch}, \& {Toomre}}]{Brun2004}
{Brun}, A.~S., {Miesch}, M.~S., \& {Toomre}, J. 2004, \apj, 614, 1073

\bibitem[{{Chandra} {et~al.}(2010){Chandra}, {Vats}, \& {Iyer}}]{Chandra2010}
{Chandra}, S., {Vats}, H.~O., \& {Iyer}, K.~N. 2010, \mnras, 407, 1108

\bibitem[{{Crooker} {et~al.}(2002){Crooker}, {Gosling}, \&
  {Kahler}}]{Crooker2002}
{Crooker}, N.~U., {Gosling}, J.~T., \& {Kahler}, S.~W. 2002, Journal of
  Geophysical Research (Space Physics), 107, 1028

\bibitem[{{Domingo} {et~al.}(1995){Domingo}, {Fleck}, \&
  {Poland}}]{Domingo1995}
{Domingo}, V., {Fleck}, B., \& {Poland}, A.~I. 1995, \solphys, 162, 1

\bibitem[{{Elsner} \& {Tsonis}(1996)}]{elsner1996}
{Elsner}, J.~B. \& {Tsonis}, A.~A. 1996, {Singular Spectrum Analysis: A New
  Tool in Time Series Analysis} (New York, NY, USA: Springer)

\bibitem[{{Fisher} \& {Sime}(1984)}]{Fisher1984}
{Fisher}, R. \& {Sime}, D.~G. 1984, \apj, 287, 959

\bibitem[{{Gardner} {et~al.}(2002){Gardner}, {Smith}, {Kohl}, {Atkins},
  {Miralles}, {Panasyuk}, {Strachan}, {Suleiman}, {Romoli}, \&
  {Fineschi}}]{Gardner2002}
{Gardner}, L.~D., {Smith}, P.~L., {Kohl}, J.~L., {et~al.} 2002, ISSI Scientific
  Reports Series, 2, 161

\bibitem[{{Geiss} {et~al.}(1970){Geiss}, {Hirt}, \& {Leutwyler}}]{geiss1970}
{Geiss}, J., {Hirt}, P., \& {Leutwyler}, H. 1970, \solphys, 12, 458

\bibitem[{{Ghil} {et~al.}(2002){Ghil}, {Allen}, {Dettinger}, {Ide},
  {Kondrashov}, {Mann}, {Robertson}, {Saunders}, {Tian}, {Varadi}, \&
  {Yiou}}]{Ghil2002}
{Ghil}, M., {Allen}, M.~R., {Dettinger}, M.~D., {et~al.} 2002, Reviews of
  Geophysics, 40, 1003

\bibitem[{{Gigolashvili} {et~al.}(2013){Gigolashvili}, {Japaridze}, \&
  {Mdzinarishvili}}]{Gigolashvili2013}
{Gigolashvili}, M.~S., {Japaridze}, D.~R., \& {Mdzinarishvili}, T.~G. 2013,
  Advances in Space Research, 52, 2122

\bibitem[{{Giordano}(1998)}]{Giordano1998t}
{Giordano}, S. 1998, PhD thesis, Univ. Torino

\bibitem[{{Giordano} \& {Mancuso}(2008)}]{Giordano2008}
{Giordano}, S. \& {Mancuso}, S. 2008, \apj, 688, 656

\bibitem[{{Groth} \& {Ghil}(2011)}]{Groth2011}
{Groth}, A. \& {Ghil}, M. 2011, \pre, 84, 036206

\bibitem[{{Hansen} {et~al.}(1969){Hansen}, {Hansen}, \& {Loomis}}]{Hansen1969}
{Hansen}, R.~T., {Hansen}, S.~F., \& {Loomis}, H.~G. 1969, \solphys, 10, 135

\bibitem[{{Hiremath} \& {Hegde}(2013)}]{Hiremath2013}
{Hiremath}, K.~M. \& {Hegde}, M. 2013, \apj, 763, 137

\bibitem[{{Hoeksema} \& {Scherrer}(1987)}]{Hoeksema1987}
{Hoeksema}, J.~T. \& {Scherrer}, P.~H. 1987, \apj, 318, 428

\bibitem[{{Horne} \& {Baliunas}(1986)}]{Horne1986}
{Horne}, J.~H. \& {Baliunas}, S.~L. 1986, \apj, 302, 757

\bibitem[{{Howe}(2009)}]{Howe2009}
{Howe}, R. 2009, Living Reviews in Solar Physics, 6, 1

\bibitem[{{Insley} {et~al.}(1995){Insley}, {Moore}, \& {Harrison}}]{Insley1995}
{Insley}, J.~E., {Moore}, V., \& {Harrison}, R.~A. 1995, \solphys, 160, 1

\bibitem[{{Karachik} {et~al.}(2006){Karachik}, {Pevtsov}, \&
  {Sattarov}}]{Karachik2006}
{Karachik}, N., {Pevtsov}, A.~A., \& {Sattarov}, I. 2006, \apj, 642, 562

\bibitem[{{Kohl} {et~al.}(1995){Kohl}, {Esser}, {Gardner}, {Habbal}, {Dennis},
  {Nystrom}, {Raymond}, {Smith}, {van Ballegooijen}, {Noci}, {Romoli},
  {Ciaravella}, {Huber}, {Antonucci}, {Giordano}, {Tondello}, {Nicolosi},
  {Pernechele}, {Spadaro}, {Poletto}, {von der L{\"u}he}, {Geiss}, {Gloeckler},
  {Allegra}, {Brusa}, {Wood}, {Siegmund}, {Fisher}, \& {Jhabvala}}]{Kohl1995}
{Kohl}, J.~L., {Esser}, R., {Gardner}, L.~D., {et~al.} 1995, \solphys, 162, 313

\bibitem[{{Kong} {et~al.}(2018){Kong}, {Pan}, {Yan}, {Wang}, \&
  {Li}}]{Kong2018}
{Kong}, D.~F., {Pan}, G.~M., {Yan}, X.~L., {Wang}, J.~C., \& {Li}, Q.~L. 2018,
  \apjl, 863, L22

\bibitem[{{Lemen} {et~al.}(2012){Lemen}, {Title}, {Akin}, {Boerner}, {Chou},
  {Drake}, {Duncan}, {Edwards}, {Friedlaender}, {Heyman}, {Hurlburt}, {Katz},
  {Kushner}, {Levay}, {Lindgren}, {Mathur}, {McFeaters}, {Mitchell}, {Rehse},
  {Schrijver}, {Springer}, {Stern}, {Tarbell}, {Wuelser}, {Wolfson}, {Yanari},
  {Bookbinder}, {Cheimets}, {Caldwell}, {Deluca}, {Gates}, {Golub}, {Park},
  {Podgorski}, {Bush}, {Scherrer}, {Gummin}, {Smith}, {Auker}, {Jerram},
  {Pool}, {Soufli}, {Windt}, {Beardsley}, {Clapp}, {Lang}, \&
  {Waltham}}]{Lemen2012}
{Lemen}, J.~R., {Title}, A.~M., {Akin}, D.~J., {et~al.} 2012, \solphys, 275, 17

\bibitem[{{Lewis} {et~al.}(1999){Lewis}, {Simnett}, {Brueckner}, {Howard},
  {Lamy}, \& {Schwenn}}]{Lewis1999}
{Lewis}, D.~J., {Simnett}, G.~M., {Brueckner}, G.~E., {et~al.} 1999, \solphys,
  184, 297

\bibitem[{{Lomb}(1976)}]{lomb1976}
{Lomb}, N.~R. 1976, \apss, 39, 447

\bibitem[{{Ma} {et~al.}(2014){Ma}, {Qu}, {Yan}, \& {Xue}}]{Ma2014}
{Ma}, L., {Qu}, Z.-Q., {Yan}, X.-L., \& {Xue}, Z.-K. 2014, Research in
  Astronomy and Astrophysics, 14, 221

\bibitem[{{Madjarska} {et~al.}(2004){Madjarska}, {Doyle}, \& {van
  Driel-Gesztelyi}}]{Madjarska2004}
{Madjarska}, M.~S., {Doyle}, J.~G., \& {van Driel-Gesztelyi}, L. 2004, \apjl,
  603, L57

\bibitem[{{Mancuso} {et~al.}(2020){Mancuso}, {Barghini}, \&
  {Telloni}}]{Mancuso2020}
{Mancuso}, S., {Barghini}, D., \& {Telloni}, D. 2020, \aap, 636, A96

\bibitem[{{Mancuso} \& {Garzelli}(2007)}]{Mancuso2007}
{Mancuso}, S. \& {Garzelli}, M.~V. 2007, \aap, 466, L5

\bibitem[{{Mancuso} \& {Giordano}(2011)}]{Mancuso2011}
{Mancuso}, S. \& {Giordano}, S. 2011, \apj, 729, 79

\bibitem[{{Mancuso} \& {Giordano}(2012)}]{Mancuso2012}
{Mancuso}, S. \& {Giordano}, S. 2012, \aap, 539, A26

\bibitem[{{Mancuso} \& {Giordano}(2013)}]{Mancuso2013}
{Mancuso}, S. \& {Giordano}, S. 2013, Journal of Advanced Research, 4, 283

\bibitem[{{Mancuso} {et~al.}(2018){Mancuso}, {Lee}, {Taricco}, \&
  {Rubinetti}}]{Mancuso2018}
{Mancuso}, S., {Lee}, T.~S., {Taricco}, C., \& {Rubinetti}, S. 2018, \solphys,
  293, 124

\bibitem[{{Mancuso} \& {Raymond}(2015)}]{Mancuso2015}
{Mancuso}, S. \& {Raymond}, J.~C. 2015, \aap, 573, A33

\bibitem[{{Mancuso} {et~al.}(2016){Mancuso}, {Raymond}, {Rubinetti}, \&
  {Taricco}}]{Mancuso2016}
{Mancuso}, S., {Raymond}, J.~C., {Rubinetti}, S., \& {Taricco}, C. 2016, \aap,
  592, L8

\bibitem[{{Nash} {et~al.}(1988){Nash}, {Sheeley}, \& {Wang}}]{Nash1988}
{Nash}, A.~G., {Sheeley}, N.~R., J., \& {Wang}, Y.~M. 1988, \solphys, 117, 359

\bibitem[{{Navarro-Peralta} \& {Sanchez-Ibarra}(1994)}]{Navarro1994}
{Navarro-Peralta}, P. \& {Sanchez-Ibarra}, A. 1994, \solphys, 153, 169

\bibitem[{{Noci} {et~al.}(1997{\natexlab{a}}){Noci}, {Kohl}, {Antonucci},
  {Tondello}, {Huber}, {Fineschi}, {Gardner}, {Naletto}, {Nicolosi}, {Raymond},
  {Romoli}, {Spadaro}, {Siegmund}, {Benna}, {Ciaravella}, {Giordano},
  {Michels}, {Modigliani}, {Panasyuk}, {Pernechele}, {Poletto}, {Smith}, \&
  {Strachan}}]{Noci1997}
{Noci}, G., {Kohl}, J.~L., {Antonucci}, E., {et~al.} 1997{\natexlab{a}},
  Advances in Space Research, 20, 2219

\bibitem[{{Noci} {et~al.}(1997{\natexlab{b}}){Noci}, {Kohl}, {Antonucci},
  {Tondello}, {Fineschi}, {Gardner}, {Nicolosi}, {Romoli}, {Maccari},
  {Raymond}, {Benna}, {Ciaravella}, {Michels}, {Modigliani}, {Panasyuk},
  {Pernechele}, {Smith}, \& {Strachan}}]{noci1997b}
{Noci}, G., {Kohl}, J.~L., {Antonucci}, E., {et~al.} 1997{\natexlab{b}}, in ESA
  Special Publication, Vol. 404, Fifth SOHO Workshop: The Corona and Solar Wind
  Near Minimum Activity, ed. A.~{Wilson}, 75

\bibitem[{{Obridko} \& {Shelting}(1989)}]{Obridko1989}
{Obridko}, V.~N. \& {Shelting}, B.~D. 1989, \solphys, 124, 73

\bibitem[{{Obridko} \& {Shelting}(2001)}]{Obridko2001}
{Obridko}, V.~N. \& {Shelting}, B.~D. 2001, \solphys, 201, 1

\bibitem[{{Parker} {et~al.}(1982){Parker}, {Hansen}, \& {Hansen}}]{Parker1982}
{Parker}, G.~D., {Hansen}, R.~T., \& {Hansen}, S.~F. 1982, \solphys, 80, 185

\bibitem[{{Pesnell} {et~al.}(2012){Pesnell}, {Thompson}, \&
  {Chamberlin}}]{Pesnell2012}
{Pesnell}, W.~D., {Thompson}, B.~J., \& {Chamberlin}, P.~C. 2012, \solphys,
  275, 3

\bibitem[{{Plaut} \& {Vautard}(1994)}]{Plaut1994}
{Plaut}, G. \& {Vautard}, R. 1994, Journal of Atmospheric Sciences, 51, 210

\bibitem[{{Ru{\v{z}}djak} {et~al.}(2017){Ru{\v{z}}djak}, {Braj{\v{s}}a},
  {Sudar}, {Skoki{\'c}}, \& {Poljan{\v{c}}i{\'c} Beljan}}]{Ruzdjak2017}
{Ru{\v{z}}djak}, D., {Braj{\v{s}}a}, R., {Sudar}, D., {Skoki{\'c}}, I., \&
  {Poljan{\v{c}}i{\'c} Beljan}, I. 2017, \solphys, 292, 179

\bibitem[{{Scargle}(1982)}]{Scargle1982}
{Scargle}, J.~D. 1982, \apj, 263, 835

\bibitem[{{Schou} {et~al.}(1998){Schou}, {Antia}, {Basu}, {Bogart}, {Chitre},
  {Christensen-Dalsgaard}, {Dziembowski}, {Eff-Darwich}, {Haber}, {Hoeksema},
  {Howe}, {Kosovichev}, {Larsen}, {Scherrer}, {Sekii}, {Title}, \&
  {Thompson}}]{Schou1998}
{Schou}, J., {Antia}, H.~M., {Basu}, S., {et~al.} 1998, \apj, 505, 390

\bibitem[{{Shelke} \& {Pande}(1985)}]{Shelke1985}
{Shelke}, R.~N. \& {Pande}, M.~C. 1985, \solphys, 95, 193

\bibitem[{{Snodgrass}(1984)}]{Snodgrass1984}
{Snodgrass}, H.~B. 1984, \solphys, 94, 13

\bibitem[{{Stix}(1989)}]{stix1989}
{Stix}, M. 1989, {The Sun. an Introduction}

\bibitem[{{Sudar} {et~al.}(2017){Sudar}, {Braj{\v{s}}a}, {Skoki{\'c}},
  {Poljan{\v{c}}i{\'c} Beljan}, \& {W{\"o}hl}}]{Sudar2017}
{Sudar}, D., {Braj{\v{s}}a}, R., {Skoki{\'c}}, I., {Poljan{\v{c}}i{\'c}
  Beljan}, I., \& {W{\"o}hl}, H. 2017, \solphys, 292, 86

\bibitem[{{S{\'y}kora} \& {Ryb{\'a}k}(2010)}]{Sykora2010}
{S{\'y}kora}, J. \& {Ryb{\'a}k}, J. 2010, \solphys, 261, 321

\bibitem[{{Taricco} {et~al.}(2015){Taricco}, {Mancuso}, {Ljungqvist},
  {Alessio}, \& {Ghil}}]{Taricco2015}
{Taricco}, C., {Mancuso}, S., {Ljungqvist}, F.~C., {Alessio}, S., \& {Ghil}, M.
  2015, Climate Dynamics, 45, 83

\bibitem[{{Temmer} {et~al.}(2006){Temmer}, {Ryb{\'a}k}, {Bend{\'\i}k},
  {Veronig}, {Vogler}, {Otruba}, \& {P{\"o}tzi}}]{Temmer2006}
{Temmer}, M., {Ryb{\'a}k}, J., {Bend{\'\i}k}, P., {et~al.} 2006, \aap, 447, 735

\bibitem[{{Timothy} {et~al.}(1975){Timothy}, {Krieger}, \&
  {Vaiana}}]{Timothy1975}
{Timothy}, A.~F., {Krieger}, A.~S., \& {Vaiana}, G.~S. 1975, \solphys, 42, 135

\bibitem[{{Ulrich} \& {Bertello}(1996)}]{ulrich1996}
{Ulrich}, R.~K. \& {Bertello}, L. 1996, \apjl, 465, L65

\bibitem[{{Uzzo} {et~al.}(2003){Uzzo}, {Ko}, {Raymond}, \& {Wurz}}]{uzzo2003}
{Uzzo}, M., {Ko}, Y.~K., {Raymond}, J.~C., \& {Wurz}, P.~and{Ipavich}, F.~M.
  2003, \apj, 585, 1062

\bibitem[{{VanderPlas}(2018)}]{VanderPlas2018}
{VanderPlas}, J.~T. 2018, \apjs, 236, 16

\bibitem[{{VanderPlas} \& {Ivezi{\'c}}(2015)}]{VanderPlas2015}
{VanderPlas}, J.~T. \& {Ivezi{\'c}}, {\v{Z}}. 2015, \apj, 812, 18

\bibitem[{{Vats} \& {Chandra}(2011)}]{Vats2011}
{Vats}, H.~O. \& {Chandra}, S. 2011, \mnras, 413, L29

\bibitem[{{Vautard} \& {Ghil}(1989)}]{Vautard1989}
{Vautard}, R. \& {Ghil}, M. 1989, Physica D Nonlinear Phenomena, 35, 395

\bibitem[{{Wang} {et~al.}(1989){Wang}, {Nash}, \& {Sheeley}}]{Wang1989}
{Wang}, Y.~M., {Nash}, A.~G., \& {Sheeley}, N.~R., J. 1989, Science, 245, 712

\bibitem[{{Wang} \& {Sheeley}(1993)}]{Wang1993}
{Wang}, Y.~M. \& {Sheeley}, N.~R., J. 1993, \apj, 414, 916

\bibitem[{{Wang} \& {Sheeley}(2004)}]{Wang2004}
{Wang}, Y.~M. \& {Sheeley}, N.~R., J. 2004, \apj, 612, 1196

\bibitem[{{Wang} {et~al.}(1988){Wang}, {Sheeley}, {Nash}, \&
  {Shampine}}]{Wang1988}
{Wang}, Y.~M., {Sheeley}, N.~R., J., {Nash}, A.~G., \& {Shampine}, L.~R. 1988,
  \apj, 327, 427

\bibitem[{{Weber} {et~al.}(1999){Weber}, {Acton}, {Alexander}, {Kubo}, \&
  {Hara}}]{Weber1999}
{Weber}, M.~A., {Acton}, L.~W., {Alexander}, D., {Kubo}, S., \& {Hara}, H.
  1999, \solphys, 189, 271

\bibitem[{{W{\"o}hl} {et~al.}(2010){W{\"o}hl}, {Braj{\v{s}}a}, {Hanslmeier}, \&
  {Gissot}}]{Wohl2010}
{W{\"o}hl}, H., {Braj{\v{s}}a}, R., {Hanslmeier}, A., \& {Gissot}, S.~F. 2010,
  \aap, 520, A29

\bibitem[{{Zaatri} {et~al.}(2009){Zaatri}, {W{\"o}hl}, {Roth}, {Corbard}, \&
  {Braj{\v{s}}a}}]{Zaatri2009}
{Zaatri}, A., {W{\"o}hl}, H., {Roth}, M., {Corbard}, T., \& {Braj{\v{s}}a}, R.
  2009, \aap, 504, 589

\bibitem[{{Zechmeister} \& {K{\"u}rster}(2009)}]{Zechmeister2009}
{Zechmeister}, M. \& {K{\"u}rster}, M. 2009, \aap, 496, 577

\end{thebibliography}

\end{document}